\documentclass[twocolumn]{aastex63}

\usepackage{graphicx}% Include figure files
\usepackage{bm}% bold math
\usepackage{hyperref}% add hypertext capabilities
\usepackage{ulem}

\begin{document}

\shorttitle{The Universe is Brighter}
\shortauthors{Darling}

\title{The Universe is Brighter in the Direction of Our Motion:  Galaxy Counts and Fluxes are Consistent with the CMB Dipole}

\author{Jeremy Darling}
 \affiliation{Center for Astrophysics and Space Astronomy \\
Department of Astrophysical and Planetary Sciences \\
University of Colorado, 389 UCB \\
Boulder, CO 80309-0389, USA}
 \email{jeremy.darling@colorado.edu}


 \begin{abstract}
   An observer moving with respect to the cosmic rest frame should observe a concentration and brightening
   of galaxies in the direction of motion and a spreading and dimming in the opposite direction.  The velocity inferred
   from this dipole should match that of the cosmic microwave background (CMB)
   temperature dipole if galaxies are on average at rest with respect to the CMB rest frame.  However, recent
   studies have claimed a many-fold enhancement of galaxy counts and flux in the direction of the
   solar motion compared to the CMB expectation, calling into question the standard cosmology.
   Here we show that the sky distribution and brightness of extragalactic radio sources are
   consistent with the CMB dipole in direction and velocity.  We use the first epoch
   of the Very Large Array Sky Survey combined with the Rapid Australian Square Kilometer Array Pathfinder
   Continuum Survey to estimate the dipole via several different methods, and all show similar results.
   Typical fits find a $331^{+161}_{-107}$~km~s$^{-1}$ velocity dipole with apex
   $(\ell,b) = (271^{+55}_{-58}, 56^{+13}_{-35})$ in Galactic coordinates from source counts and
   $399^{+264}_{-199}$~km~s$^{-1}$ toward $(\ell,b) = (301^{+30}_{-30}, 43^{+19}_{-17})$
   from radio fluxes.  These are consistent with the CMB-solar velocity, 370~km~s$^{-1}$ toward
   $(\ell,b) = (264, 48)$, and show
   that galaxies are on average at rest with respect to the rest frame of the early universe,
   as predicted by the canonical cosmology.  
\end{abstract}

\section{\label{sec:intro}Introduction}

The cosmic microwave background (CMB) shows a 3.4~mK dipole that is interpreted as a peculiar
Solar motion with respect to the CMB rest frame \citep{smoot1977}.
If this interpretation is correct, then our canonical cosmology
predicts that galaxies should show a corresponding dipole in apparent number density and brightness
\citep{ellis1984}.  Galaxies should also show a secular parallax dipole in their proper motions
\citep{ding2009, paine2020, croft2021}.
Detection of a galaxy number, flux, or parallax dipole that is in agreement with the CMB dipole would confirm the
usual interpretation of the
CMB dipole and indicate that the galaxy sample is on average at rest with respect to the CMB rest frame.

Previous work using radio galaxy counts and/or flux has produced mixed and contradictory results,
often using the same data sources, including the 1.4~GHz NRAO VLA Sky Survey \citep[NVSS;][]{condon1998},
the 843~MHz Sydney University Molonglo Sky Survey \citep[SUMSS;][]{murphy2007},  the 325~MHz
Westerbork Northern Sky Survey \citep[WENSS;][]{rengelink1997}, and the 147~MHz TIFR GMRT Sky Survey
\citep[TGSS-ADR1;][]{intema2017}.  \citet{blake2002} found NVSS counts that were consistent with the CMB dipole.
Subsequent studies, including those based on the NVSS, found rough agreement with the CMB dipole direction but
higher than expected dipole amplitude, of order 1--6\% variation rather than the canonical $\sim0.5$\% amplitude,
and therefore excluded the CMB-Solar velocity with high confidence \citep[e.g.,][]{singal2011,rubart2013,tiwari2019,siewert2021}.
\citet{gibelyou2012} found significant disagreement with the CMB in direction and amplitude.
In infrared bands, \citet{secrest2021} used a quasar catalog selected from the
Wide-field Infrared Survey Explorer \citep[WISE;][]{wright2010} to reject the kinematic interpretation of the CMB
dipole at 4.9$\sigma$ significance.  Other infrared work focused on low-redshift galaxies that may be significantly
influenced by large scale structure \citep[e.g.,][]{gibelyou2012}.

Consistency of extragalactic objects with the CMB dipole has been found in galaxy redshift surveys
\citep[e.g.,][]{rowan-robinson1990,lavaux2010},
  in the modulation of the thermal Sunyaev-Zeldovich effect \citep[tSZ;][]{akrami2020} and in supernovae \citep{horstmann2021},
suggesting that the discrepancy lies with galaxy and quasar surveys rather than with the canonical cosmology, large-scale structures, or lensing, as some authors have suggested.

In this work, we employ two new radio continuum surveys, the  Karl G. Jansky Very Large Array\footnote{The National Radio Astronomy Observatory is a facility of the National Science Foundation operated under cooperative agreement by Associated Universities, Inc.} Sky Survey \citep[VLASS;][]{vlass2020}
in the northern hemisphere and the Rapid Australian Square Kilometer Array Pathfinder\footnote{The Australian SKA Pathfinder is part of the Australia Telescope National Facility which is managed by CSIRO.}
Continuum Survey \citep[RACS;][]{racsI_2020} in the south, to measure the galaxy number and flux dipoles
using nearly complete (90\%) sky coverage.  
We present the theoretical expectations for galaxy dipoles in number and flux in Section \ref{sec:theory}, we
discuss the data sources, catalog-matching, and map-making in Section \ref{sec:data}, and we describe
four different dipole measurement techniques for number and flux in Section \ref{sec:meas}.   The results of
these measurements are all self-consistent and in good agreement with the CMB dipole (Section \ref{sec:results}), and
it is not clear why this work produced such different results than previous studies (Section \ref{sec:discussion}).
We conclude with some suggestions for future improvement on this analysis, both in terms of technique and looking
to future data from ongoing and planned continuum surveys (Section \ref{sec:conclusions}).  

For comparison to the results in this work, we use the fiducial
\citet{planck2020_I,planck2020_III} CMB dipole:  $3362.08\pm0.99$~$\mu$K toward galactic coordinates $(\ell,b) = (264\fdg021\pm0\fdg011, 48\fdg253\pm0\fdg005)$ or equatorial coordinates $(\alpha,\delta) = (167\fdg942 \pm 0\fdg007,-6\fdg944\pm0\fdg007)$ (J2000).  Absent an intrinsic CMB dipole, the velocity of the Sun relative to the CMB is $369.82\pm 0.11$~km~s$^{-1}$.

\begin{figure*}[t]
\begin{centering}
  \includegraphics[scale=0.22,trim= 0 60 0 60,clip=true]{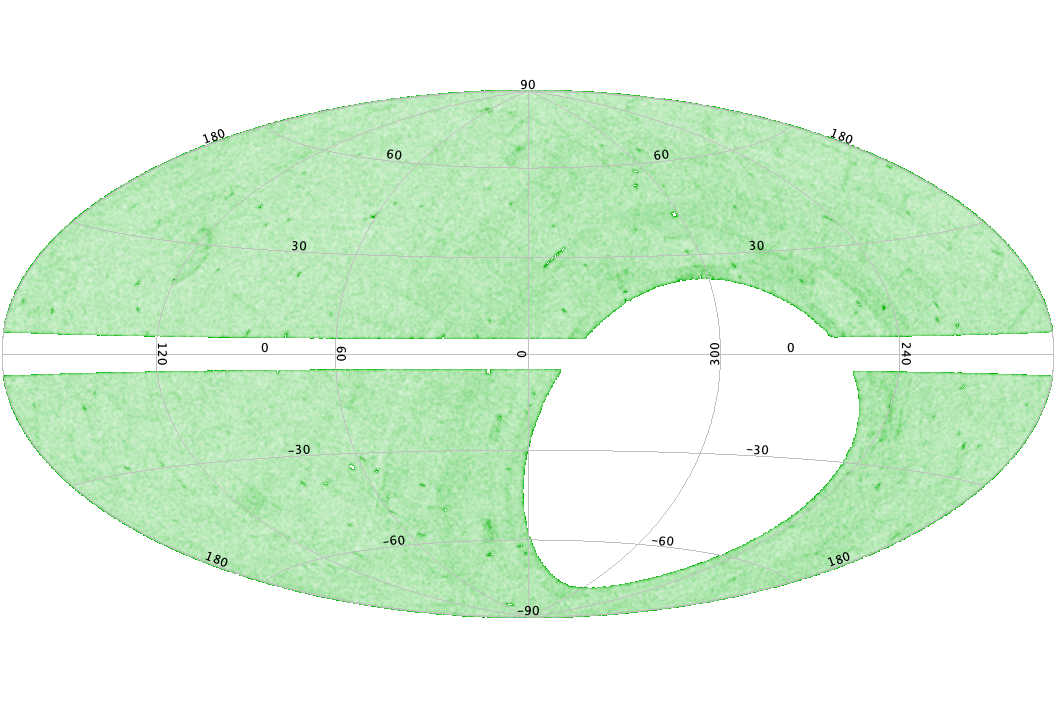}
  \includegraphics[scale=0.22,trim= -50 60 0 60,clip=true]{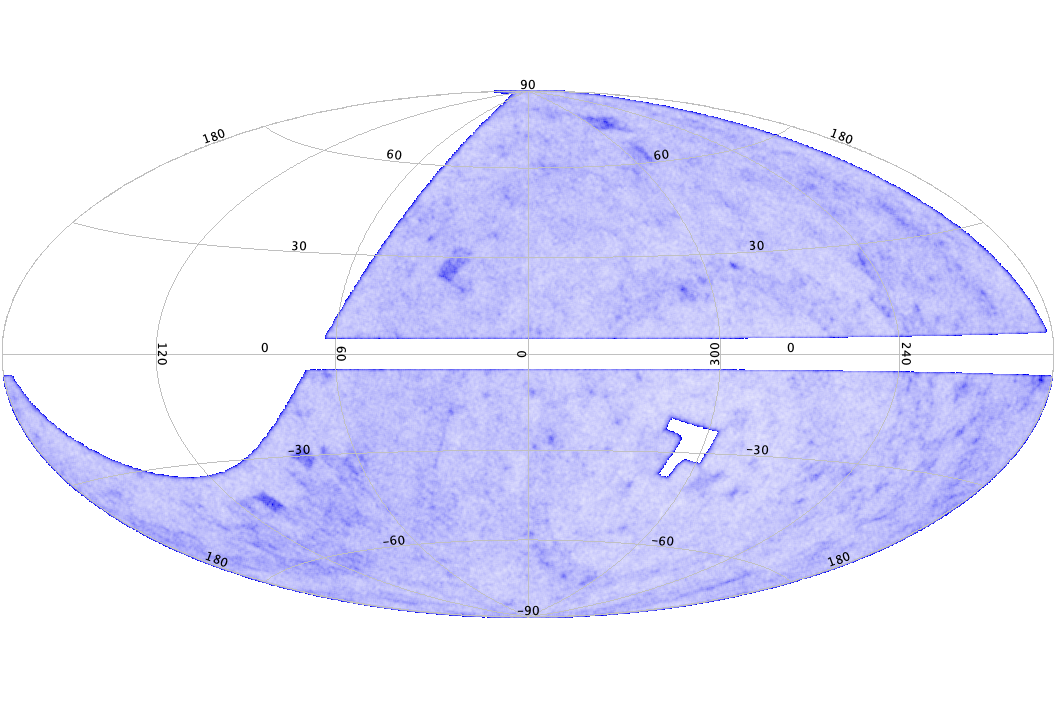}
\caption{  VLASS (left) and RACS (right) catalog sources in galactic coordinates.  Darker color indicates fewer sources.
  $|b|<5^\circ$ was excluded from the VLASS.
}
\label{fig:surveys}
\end{centering}
\end{figure*}

\section{\label{sec:theory} Theoretical Expectations}

\citet{ellis1984} showed how the number density of galaxies would be altered for an observer moving with
respect to the CMB rest frame \citep[see also][]{gibelyou2012,rubart2013}.
Briefly, starting with
Lorentz-invariant quantities such as number $N$ and a frequency-normalized specific intensity, $I_\nu/\nu^3$,
which is the photon phase space distribution, one can show that an observer Lorentz-boosted 
with respect to the average rest frame of galaxies will observe a dipole on the sky
that depends on powers of the factor $\delta$, 
\begin{equation}
  \delta = \gamma (1+\hat{n}\cdot\vec{v}) = {1+v \cos \theta \over \left(1-v^2\right)^{1/2}}\ ,
\end{equation}
for an observation in direction $\hat{n}$ and boost $\vec{v}$ ($c \equiv 1$).
The observed effects are:
(1) a relativistic Doppler shift, $\delta^1$; 
(2) aberration, which alters solid angles as $\delta^2$; 
(3) flux modification of objects with power-law spectral energy distributions $S_\nu \propto \nu^{-\alpha}$
when observed at fixed frequencies, $\delta^{1+\alpha}$; and 
(4) a change in the number of galaxies detected using a fixed flux limit, if $N(>S) \propto S^{-x}$, $\delta^{x(1+\alpha)}$.
Combining these effects to first order in $v$, we expect the
sky surface density of galaxies will be modulated by the angular distance $\theta$ from the dipole apex
\begin{equation}
  {dN \over d\Omega} = \left.{dN \over d\Omega}\right|_{\theta=\pi/2}\left(1+ [2+x (1+\alpha)]\, v \cos\theta\right)\ .
\end{equation}
The expected fractional change in a monopole-subtracted sky would therefore be
\begin{equation}\label{eqn:DN}
  {\Delta N \over N_{\ell=0}} = [2+x (1+\alpha)]\, v \cos\theta\ .
\end{equation}
The integrated flux density per solid angle brings an extra factor of $\delta^{1+\alpha}$, and the fractional change becomes
\begin{equation}\label{eqn:DS}
  {\Delta S \over S_{\ell=0}} = [3+\alpha+x (1+\alpha)]\, v \cos\theta\ .
\end{equation}
This is the same quantity as the flux-weighted galaxy count \citep{rubart2013}.
Turning Equations \ref{eqn:DN} and \ref{eqn:DS} around, for an observed null-centered (monopole-subtracted) dipole amplitude on the sky,
$\mathcal{D}_N$ or $\mathcal{D}_S$, the observer's velocity with respect to a galaxy catalog can be calculated:
\begin{eqnarray}
 v & = &  \mathcal{D}_N / [2+x (1+\alpha)]  \ \ {\rm or} \label{eqn:vN}\\
 v & = &  \mathcal{D}_S / [3+\alpha+x (1+\alpha)]\ . \label{eqn:vS}
\end{eqnarray}
If the galaxy catalog is at rest with respect to the CMB, then the measured velocity should match the CMB velocity
dipole in direction and amplitude.  The direction of the dipole is obtained from the apex coordinates.

In principle, the flux dipole should be more sensitive than number because higher signal-to-noise can be obtained
from flux measurements.  But in practice, uniform flux calibration across a large survey (or multiple surveys) is
challenging, and constraining systematics well below 1\% is difficult.  Simply counting well-detected sources
is less challenging if one works above the survey completeness limit (although multi-component radio sources
can complicate the counting exercise).

\section{\label{sec:data} Data}

Sky dipole measurements require all-sky coverage, which can be done using space-based
facilities such as WISE \citep{wright2010,secrest2021}, but require dual-hemisphere terrestrial surveys.
For this work, we use two new radio continuum surveys:  the VLASS at 3~GHz in the north \citep{vlass2020}
and the RACS at 887.5~MHz in the south \citep{racsI_2020}.

The VLASS is a three-epoch synoptic survey spanning 2--4~GHz at declinations $\delta > -$40$^\circ$ with
2.5\arcsec\ angular resolution.  The
first epoch data release catalog includes $1.9\times10^6$ sources with 99.8\% completeness at 3 mJy \citep{gordon2021}.
We select \texttt{Duplicate\_flag} $< 2$,
\texttt{Quality\_flag == 0}, and $|b| > 5^\circ$ from the Component Table (version 2) to obtain $\sim$1.4 million sources.
The first VLASS epoch was split (epochs 1.1 and 1.2), and the two sub-epochs show
different noise properties and systematic flux density offsets compared to expectations \citep{gordon2021}.
Nevertheless, the $4^\circ\times10^\circ$ survey tiling was
interleaved somewhat\footnote{Kimball, A. 2017, VLASS Project Memo \#7: VLASS Tiling and Sky
  Coverage, \url{https://library.nrao.edu/public/memos/vla/vlass/VLASS_007.pdf}} on angular scales much smaller
than the $\ell=1$ dipole scale, and our analysis does not include sources close to the detection limit or the portion
of the survey that uses the hybrid array (see below).
Following the advice of \citet{gordon2021}, all flux densities are scaled by 1/0.87.  

The RACS first data release catalog consists of $\sim$2.1 million radio sources in declinations
$-$80$^\circ < \delta < +$30$^\circ$ and $|b| > 5^\circ$ with 95\% point source completeness
at $\sim$3~mJy \citep{racsII_2021}, more than a factor of three below our adopted lower flux cutoff.
This ``RACS-low'' release has a 288~MHz bandwidth centered at 887.5~MHz.  The
survey shows spatially-varying median rms noise structure between 0.1 and 0.6 mJy~beam$^{-1}$ on
$\sim$5$^\circ$ scales and shows generally increasing noise with declination.  
The RACS survey was smoothed to a uniform 25\arcsec\ resolution.

All subsequent analysis uses the peak flux densities %(mJy~beam$^{-1}$)
from these survey catalogs rather than integrated flux
densities\footnote{We use the shorthand ``flux'' in subsequent discussion to generically refer to summed flux densities in map pixels and the peak flux densities of individual galaxies.}.
Peak flux densities are equivalent to integrated flux densities for unresolved radio sources and mitigate somewhat
the angular scale and frequency mismatch between the VLASS and the RACS (see Fig.\ 9 in \citealt{gordon2021} for an example of
the relationship between peak and total flux density).
The size of radio sources is a function of observed frequency and observing configuration, which complicates
an all-sky analysis of source counts and flux using disparate surveys.

\begin{figure}[t]
\begin{centering}
  \includegraphics[scale=0.34,trim= 20 20 0 0,clip=false]{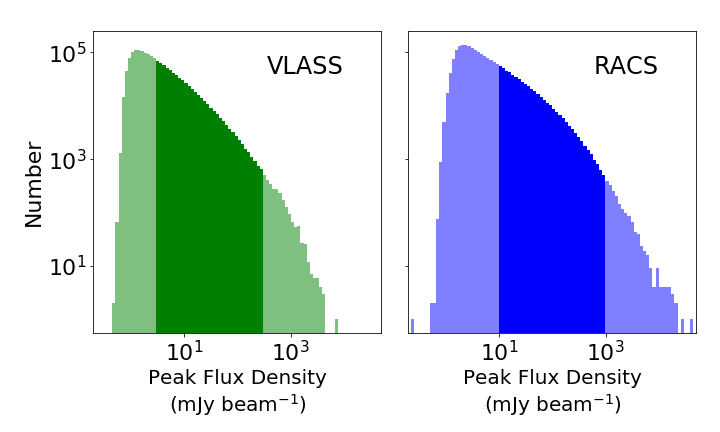}
\caption{  VLASS (left) and RACS (right) catalog source count versus peak flux density.  The opaque color indicates the
  catalog selection regions prior to masking and trimming.
}
\label{fig:counts}
\end{centering}
\end{figure}

Figure \ref{fig:surveys} shows the two survey catalogs in Galactic coordinates before they are flux-selected,
trimmed, masked, or combined.
Figure \ref{fig:counts} shows the number of sources in each catalog versus peak flux density prior to masking and trimming the samples.
We select the VLASS 3--300~mJy beam$^{-1}$ flux range\footnote{An upper flux cutoff does not alter the galaxy count or combined flux dependence on the $\delta$ factor.} based on the linear part of the $\log N$--$\log S$ plot, staying well above the
flux limit turnover at 1.3~mJy, which also reduces completeness issues caused by areally- and temporally-variable noise limits
below the 3~mJy completeness limit, and
well below the distribution turnover around 500~mJy.  This selection also avoids stochastic contributions from from bright
radio sources.  Using the spectral indices described in Sec.\ \ref{subsec:matching} below, we transfer the VLASS flux limits to the
RACS catalog.  The RACS limits for number counts and flux differ slightly but are well above the RACS 3~mJy survey completeness limit.

For the purposes of this analysis, absolute flux calibration is unimportant.  Relative flux calibration across the sky does
impact this study, but only if there are flux calibration systematics in on steradian scales.  The two-sub-epoch tiling pattern
of the VLASS does not align with the CMB dipole and could not produce a spurious coincidentally-aligned signal.

\subsection{Catalog Matching}\label{subsec:matching}

The VLASS shows elevated noise at $\delta > 80^\circ$ and the RACS is incomplete at
$\delta < -80^\circ$, so we mask both surveys
to exclude sources within $10^\circ$ of the equatorial poles.
We cut and combine the two surveys at $\delta = 0^\circ$, which excludes each survey's lowest-elevation
declination range that tends to show higher noise and instrumental artifacts (Figure \ref{fig:surveys}).
We also mask the Galactic plane for $|b| < 5^\circ$.  Excluding a larger latitude range had little impact on results,
and minimal masking reduces systematics and enhances the sky coverage \citep[e.g.,][]{siewert2021}.
We did no additional masking. The final maps cover a high sky fraction, $f_{\rm sky} = 0.90$, and
dipole measurements are therefore not expected to be significantly affected in amplitude or uncertainty by coupling to higher-order modes such as quadrupole or octopole \citep{gibelyou2012}.

For arbitrary flux limits, the two catalogs have differing source density and integrated flux density.  This
produces a dipole aligned with the equatorial poles.  Since the expected signal is of order 0.5\%, the catalog
matching is crucial to the measurement and is extremely sensitive to the spectral index used to scale the RACS
catalog at 887.5~MHz to the VLASS 3~GHz fluxes.  

For any given 3 GHz flux density, the 3.0--0.9~GHz spectral index spans $\alpha = 1\pm2$, so any flux scaling to equalize
the two surveys is only correct in aggregate and the total scaled flux is sensitive to the choice of spectral index.
Incorrect scaling leads to a hemispherical dipole (which is, in part, the reason we chose an equal-area division
between the two surveys).
We determine the median spectral index using the overlap region between the two surveys that is least affected by
noise systematics, $-10^\circ < \delta < +10^\circ$ and $|b| > 5^\circ$.  The median spectral index is $\alpha = 0.98$,
which we use to scale the flux limits of RACS to match those of the VLASS for the source number density analysis.  
For the flux analysis, we use the median RACS flux-weighted spectral index of $\alpha = 1.02$.  These spectral indices
are steeper than, but within the variance of, the 3--1.4~GHz and 1.4--0.9~GHz spectral indices \citep{racsII_2021,gordon2021}.

The combined number catalog contains 711,450 sources, with a division of 49.9\% to 50.1\%, VLASS to RACS.  The
combined flux catalog contains 697,753 sources, divided 50.9--49.1.  The solid angle ratio of the unmasked portions
of VLASS to RACS is 1.0006.
The higher resolution of VLASS compared to RACS suggests that the VLASS would detect more objects per
solid angle, but perhaps surprisingly, the areal density of sources in the surveys, after scaling
flux limits, is the same to within 0.2\%, and a mismatch does not seem to obtain for the range shown in Figure \ref{fig:counts}.

The source count $N(>S) \propto S^{-x}$ index $x$ can be determined empirically from the
catalog: we find $x=1.0$ for VLASS and RACS in the lower decade of the selected
flux range, which dominates the source counts.  The expected dipole amplitudes,
based on the above $\alpha$ values and Equations \ref{eqn:DN} and \ref{eqn:DS}, are therefore 
$\mathcal{D}_N = 4.0\, v = 0.0050$ for number and $\mathcal{D}_S = 6.1\, v = 0.0075$ for flux,
assuming a null intrinsic CMB dipole (i.e., the dipole amplitude is solely caused by the barycenter motion with
respect to the CMB rest frame). The observed dipole signals will therefore amount to less than 1\% of the mean (monopole) count or flux value.

\subsection{Map-Making}

\begin{figure*}[t]
\begin{centering}
  \includegraphics[scale=0.4,trim=0 20 0 0,clip=false]{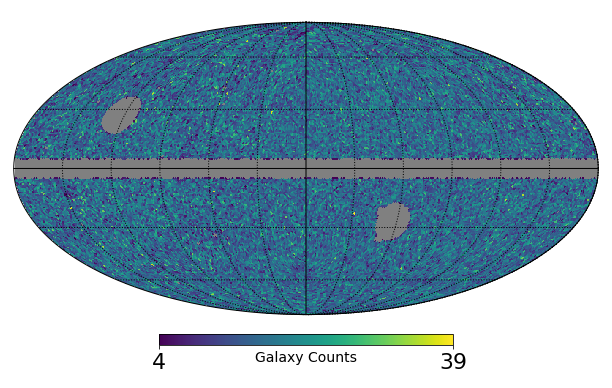}
  \includegraphics[scale=0.4,trim=0 20 0 0,clip=false]{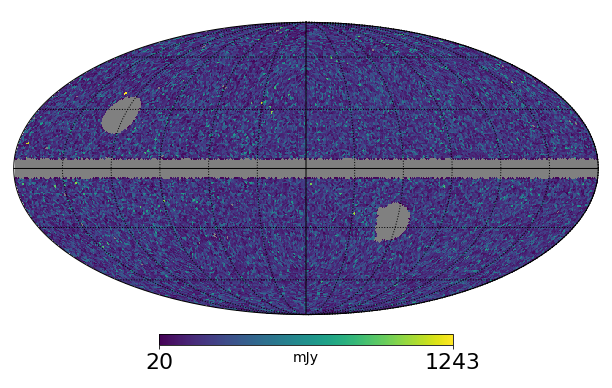}
  \includegraphics[scale=0.4,trim=0 20 0 0,clip=false]{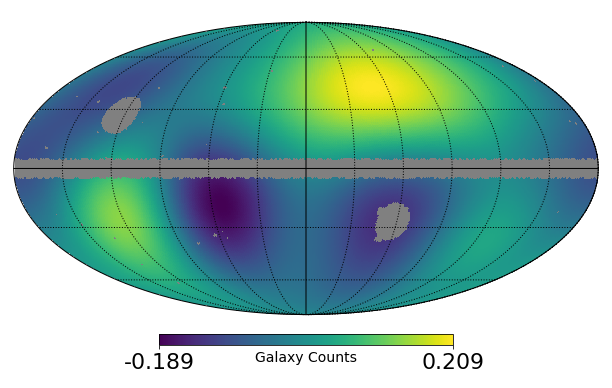}
  \includegraphics[scale=0.4,trim=0 20 0 0,clip=false]{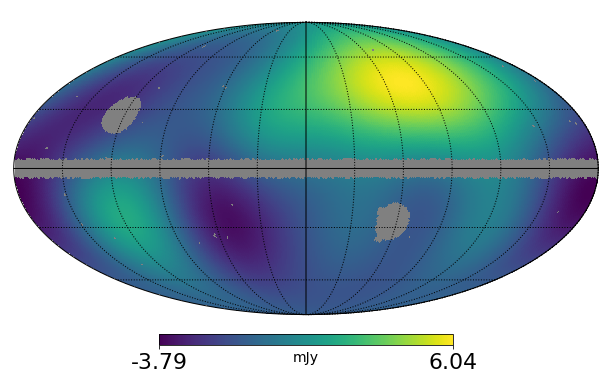}
  \includegraphics[scale=0.4,trim=0 20 0 0,clip=false]{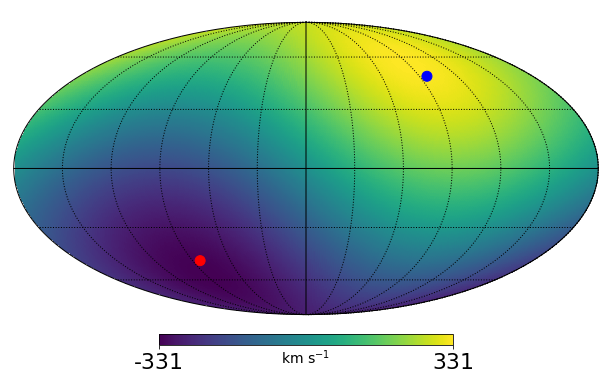}
  \includegraphics[scale=0.4,trim=0 20 0 0,clip=false]{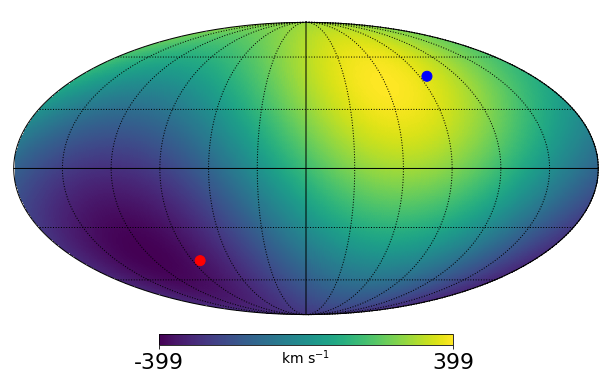}
\caption{
  Top:  HEALPix maps of number counts (left) and integrated flux (right) for the scaled and combined surveys in galactic coordinates.  The grey regions are masked.  
  Middle: Number (left) and flux (right) mean-subtracted maps smoothed to 1 ster.  These maps should not
  be confused with fractional difference maps, which are unitless.
  Bottom:   Spherical harmonic fit velocity maps for number (left) and flux (right).
  The blue and red points indicate the apex and nadir of the CMB dipole \citep{planck2020_I,planck2020_III}.
}
\label{fig:maps}
\end{centering}
\end{figure*}

To make number and flux maps we create HEALPix\footnote{\url{http://healpix.sf.net}; \citet{healpix2005}}
maps using \texttt{healpy}\footnote{\citet{healpy2019}} with much finer resolution than the expected signal in order to create minimal-area masks.  Mask edge pixels, which incompletely sample the sky, are trimmed.  
Using \texttt{nside = 64} gives 49,152 pixels with solid angle of 0.84 square degrees per pixel.
The median unmasked pixel contains 16 radio sources and 202 mJy (scaled to 3~GHz on the flux-corrected VLASS scale).
Figure \ref{fig:maps} (top) shows the mean-subtracted masked HEALPix number and flux maps.  

Maps smoothed to 1 ster show structure at the 1--3\% level (Figure \ref{fig:maps}, middle), including a
maximum in the vicinity of the CMB apex.  However, all measurements described below use the unsmoothed masked maps
or the combined catalog.

\section{\label{sec:meas} Measurements}

We estimate the number and flux density dipoles using four methods:
(1) by computing spherical harmonics (SH) components directly from the merged catalog \citep[after][]{blake2002};
(2) by fitting SH to masked sky maps using the \texttt{healpy.anafast} algorithm;
(3) using the dipole vector estimator employed by \citet{secrest2021} on maps, \texttt{healpy.fit\_dipole}; and
(4) using a ``permissive'' fit of a SH dipole model to the maps that allows for outliers \citep{sivia2006,darling2018}.

(1) The direct method does not rely on making maps.  Rather, it employs the standard method to
calculate SH coefficients $a_{\ell,m}$ from $N$ objects at $(\theta_i,\phi_i)$:
\begin{equation}
  a_{\ell,m} = \sum_{i=1}^N f_i\, Y^*_{\ell,m}(\theta_i,\phi_i)
\end{equation}
where $f_i = 1$ for number counts, and $f_i = S_i$ for flux.  For real $f_i$, the dipole map of
a quantity becomes
\begin{equation}
  f(\theta,\phi) =   a_{1,0} Y_{1,0}(\theta,\phi) +2\, a_{1,1}^\Re Y_{1,1}^\Re(\theta,\phi)
                                                                       -2\, a_{1,1}^\Im Y_{1,1}^\Im(\theta,\phi).
\end{equation}
The amplitude of the dipole can be obtained by scaling the number $N$ or flux $S$ map maximum by the monopole:
\begin{equation}
  \mathcal{D}_{N,S} = {f_{N,S}^{\rm max} \over a_{0,0}}\, \sqrt{4\pi\over3}\ .
\end{equation}
Uncertainties in the dipole amplitude and direction are estimated by
re-calculating the dipole from a 10,000-iteration bootstrap resampling of the source catalog.

(2) The \texttt{anafast} SH method fits monopole and dipole SH coefficients to masked maps rather
than datapoints.  Masked regions are omitted from the fits, and data have been aggregated into equal-area
pixels (either source counts or summed fluxes).  Uncertainties are estimated from a non-parametric bootstrap (bootstrap
resampling of fit residuals rather than datapoints).
The parametric bootstrap uncertainties agree with the non-parametric implementation for the number maps but cannot
be done on the flux maps.

(3) The dipole vector estimator described by \citet{secrest2021} uses \texttt{healpy.fit\_dipole} to
simultaneously fit a monopole $\mu$ and a Cartesian 3D dipole vector $\vec{d}$ to maps.
The fractional amplitude of the dipole signal is 
\begin{equation}
  \mathcal{D}_{N,S} = {|\vec{d}_{N,S}|\over \mu}\ ,
\end{equation}
and $\vec{d}$ points in the direction of the apex.  
We used a non-parametric bootstrap to estimate uncertainties.

(4)  This method employs a ``permissive'' fit of an SH dipole to maps
that does not assume Gaussian error distributions and thus
allows for data outliers (such as clustered radio sources).  We maximize the likelihood associated with a
probability density function based on the data-model residual $R_i$ of each pixel:
\begin{equation}
  {\rm prob}(R_i) \propto \left( 1-e^{-R_i^2/2} \over R_i^2 \right)
\end{equation}
\citep{sivia2006, darling2018}.  We use the Python package \texttt{lmfit}\footnote{\citet{newville2021}} to obtain least-squared
fits of SH dipole coefficients and their marginalized uncertainties. 
We then use a Monte Carlo realization of those error budgets to determine the dipole velocity parameters and
uncertainties (amplitude and direction).

\section{\label{sec:results}Results}

To calculate velocities from dipole amplitudes using Equations \ref{eqn:vN} and \ref{eqn:vS},
we use the spectral index $\alpha = 0.98$ for number, $\alpha = 1.02$ for flux, 
and source count index $x=1.0$ for both (Section \ref{subsec:matching}).   
Uncertainties are calculated using percentiles and are not always symmetric.
We bias-correct the results when appropriate, including rescaling the 1$\sigma$
confidence intervals (\citealt{efron1994}; note that jackknife-calculated ``accelerations'' are insignificant).  Dipole directions show negligible bias and are not corrected, but the dipole amplitudes do show bias, particularly the
dipole vector and permissive fit methods (methods 3 and 4).

\begin{deluxetable}{clcccl}
  \tablecaption{\label{tab:results} Radio Galaxy Dipole Results}
  \tablehead{  \# & \colhead{Method} & \colhead{Quantity} & \multicolumn{2}{c}{Apex} &  \colhead{Velocity} \\[-4pt] \cline{4-5} 
  & & & \colhead{$\ell$  ($^\circ$)} & \colhead{$b$  ($^\circ$)} & (km~s$^{-1}$)} %\\[-3pt]}
%  & & & ($^\circ$) & (km~s$^{-1}$) }
  \startdata
 1 & Direct & Number  & $265^{+47}_{-44}$ & $59^{+11}_{-29}$ & $225^{+116}_{-51}$  \\
            & & Flux & $300^{+24}_{-23}$ & $45^{+13}_{-17}$ & $422^{+147}_{-83}$ \\
 2 &    SH & Number  & $271^{+55}_{-58}$ & $56^{+13}_{-35}$ & $331^{+161}_{-107}$\\
           & & Flux & $301^{+30}_{-30}$ & $43^{+19}_{-17}$ & $399^{+264}_{-199}$\tablenotemark{*} \\
 3 &   $\vec{d}$ & Number & $271^{+51}_{-52}$ & $52^{+13}_{-34}$ & $259^{+152}_{-124}$\tablenotemark{*} \\
       & & Flux & $301^{+25}_{-25}$ & $39^{+15}_{-20}$ & $250^{+320}_{-160}$\tablenotemark{*} \\
  4 & Permissive & Number & $259^{+47}_{-41}$ & $50^{+12}_{-35}$ & $297^{+150}_{-120}$\tablenotemark{*} \\
      & & Flux & $299^{+23}_{-27}$ & $40^{+14}_{-19}$ & $622^{+211}_{-200}$\tablenotemark{*} \\
  \hline
  & CMB & Temperature & 264.0 & 48.3 & 369.8 
\enddata
\tablenotetext{*}{Bias-corrected velocities.}
\end{deluxetable}

\begin{figure*}[t]
\begin{centering}
  \includegraphics[scale=0.46,trim=10 14 20 30,clip=true]{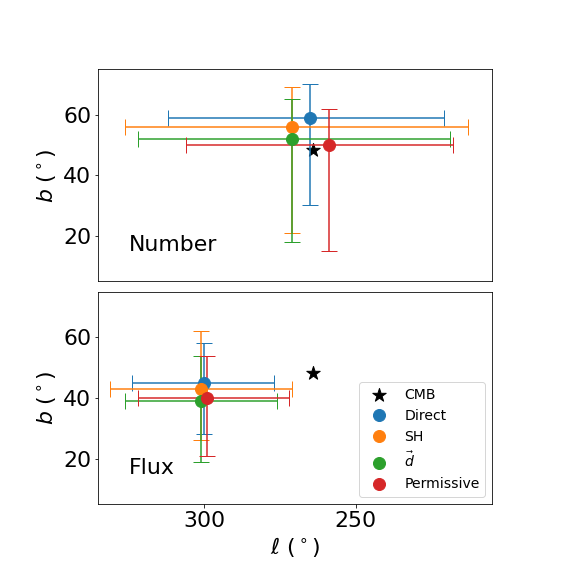}
  \includegraphics[scale=0.46,trim=10 14 20 30,clip=true]{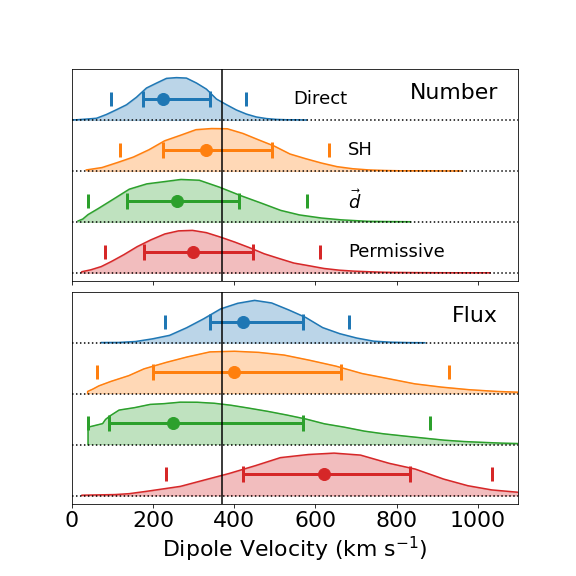}
\caption{
  Left:  Dipole apex measured from radio source number counts (top) and fluxes (bottom) using the methods
  listed in Table \ref{tab:results}.  The star indicates the CMB dipole apex, and error bars are 1$\sigma$ uncertainties.
  Right:  Dipole velocity measured from radio source number counts (top) and fluxes (bottom).
  The vertical line indicates the observed CMB dipole.  The envelopes show the error distributions in dipole amplitudes, and 
  the vertical tick marks show 1$\sigma$ and 2$\sigma$ uncertainties.
}
\label{fig:fits}
\end{centering}
\end{figure*}

Table \ref{tab:results} shows the dipole apex and velocity of the four fitting methods
for number and flux.  Figure \ref{fig:fits} shows that
all values are consistent with the CMB dipole vector (direction and velocity).
  A comparison of each dipole vector to the CMB yields $p$-values $\geq 0.64$ for number and $\geq 0.35$ for flux based on a $\chi^2$ test with 3 degrees of freedom.
Surprisingly, the permissive fit method, which is intended to omit flux outliers and clustered
sources, shows a higher flux dipole amplitude than the other methods (although it is consistent with these
and the CMB).
Figure \ref{fig:maps} (bottom) shows the velocity dipole maps for number and flux for the SH fit method.

\section{\label{sec:discussion}Discussion}

Why do these results differ from previous studies that reject the CMB dipole amplitude?
As usual, when astronomical observations are used to obtain sub-percent measurements, systematics
are a concern, and there are differences in approach and data sources to consider.  
Sky completeness is known to impose a strong bias on results \citep[e.g.,][]{gibelyou2012,siewert2021}, and this
work is among the most complete with $f_{\rm sky} = 0.90$.  The source density is comparable to previous work
\citep[see e.g.,][]{siewert2021}, the angular resolution is higher in the VLASS but not the RACS, and the
sensitivity is better at comparable frequencies.
Our results have larger uncertainties on the dipole direction than most previous work, but this is likely to be
due in part to the smaller amplitude, given the only slightly larger source catalog.
Also, previous work uses integrated fluxes, which are sensitive to source sizes, resolution, and surface
brightness sensitivity in a frequency-dependent manner.  This was the motivation for employing peak fluxes instead.  

Intriguingly, \citet{siewert2021} found that the measured dipole amplitude obtained from source counts
decreases with increasing frequency, up to 1.4~GHz (NVSS).  However, an extrapolation of their fit of $\mathcal{D}_N$
vs.\ $\nu$ predicts a value at 3~GHz of more than twice what we measure.  Nevertheless, it could be that
the Solar peculiar motion is best measured at higher radio frequencies.

While the results of this study are consistent with the CMB dipole vector, they are not necessarily inconsistent with
  previous work because the amplitude uncertainty distributions have large high-velocity tails (Figure \ref{fig:fits}).
  For example, the vector method ($\vec{d}$) for number density that is most directly comparable to
  the \citet{secrest2021} methods and result has a +3$\sigma$ value of 740~km~s$^{-1}$, which
  is roughly consistent with the \citet{secrest2021} amplitude measurement.  

The Appendix presents an empirical examination of our analysis assumptions and possible measurement systematics.
  We assess the elevation-dependent source counts in the surveys, the ability of each survey alone to measure the dipole,
  the impact of the selected flux limit, the choice of peak versus total flux, and the fine-turning of the spectral
  index used to combine the two surveys.  Nearly all of these tests recover dipoles consistent with the CMB, albeit with
  larger uncertainties in the dipole amplitude and direction than those listed in Table \ref{tab:results}.

\section{\label{sec:conclusions}Conclusions}

Contrary to previous studies, this work shows that the observed dipole in extragalactic source counts and fluxes
is consistent with an observer-induced CMB dipole; there is no discrepancy in direction or amplitude
between radio sources and the CMB.   This aligns with already CMB-consistent observations of local large scale structure, the tSZ effect, and supernovae \citep{rowan-robinson1990,lavaux2010,akrami2020,horstmann2021}.
It seems plausible that future all-sky extragalactic surveys may achieve the
depth and fidelity to either confirm that the CMB dipole is nearly all obsever-induced or to identify a contribution that
is cosmological.  

A number of refinements in the above methods can be explored, including employing a broken or two-index power law
for source counts and spectral indices following \citet{siewert2021},
using survey depth- and artifact-weighted maps, or the more formal (and perhaps more rigorous) analyses
employed by previous investigators.  
With the additional VLASS epochs and the ASKAP follow-on to RACS, the surveys themselves
will also improve soon, both in depth, calibration, and in processing refinement.

\acknowledgments
We thank the scientific, operations, observing, and computing staff at the NRAO and the ATNF
  who made this work possible and the data publicly available.
  We also thank the anonymous referee for excellent and thoughtful feedback.
  This scientific work uses data obtained from the Murchison Radio-astronomy Observatory, and we acknowledge
  the Wajarri Yamatji people as the traditional owners of the Observatory site.
  Some of the results in this paper have been derived using the HEALPix and healpy packages \citep{healpix2005,healpy2019} and TopCAT \citep{topcat2005}.
  This research made use of NumPy \citep{NumPy}, Matplotlib \citep{Matplotlib}, and
  Astropy\footnote{http://www.astropy.org}, a community-developed core Python package for Astronomy \citep{astropy:2013, astropy:2018}.

  \facility{VLA, ASKAP}
  \software{astropy \citep{astropy:2013, astropy:2018}, NumPy \citep{NumPy}, Matplotlib \citep{Matplotlib}, TopCAT \citep{topcat2005}, HealPix \citep{healpix2005}, healpy \citep{healpy2019}}

%\vfill  
%  \clearpage
  
 \bibliography{ms} % Produces the bibliography via BibTeX.

%\clearpage

\appendix

Here we examine assumptions used in the data selection and analysis, particularly regarding the combination of the two radio surveys.
The goal is twofold:
(1) to build confidence in the dipole measurement methods in light of the disparate prior results, and
(2) to explore possible reasons for discrepancies.  

Since the dipoles obtained from galaxy counts and flux are similar for all different fitting methods (Table \ref{tab:results}),
the following tests focus on galaxy counts and use the permissive least-squares fitting method.  Unless otherwise specified,
all selection parameters and methods are unchanged from Section \ref{sec:meas}.  For example, when we examine the two
radio continuum surveys individually, the analysis, flux limits, and other paramenters are unchanged unless explicitly described.

\section{Single-Survey Tests}

\begin{figure}[h]
\begin{centering}
  \includegraphics[scale=0.52,trim= 20 20 0 0,clip=false]{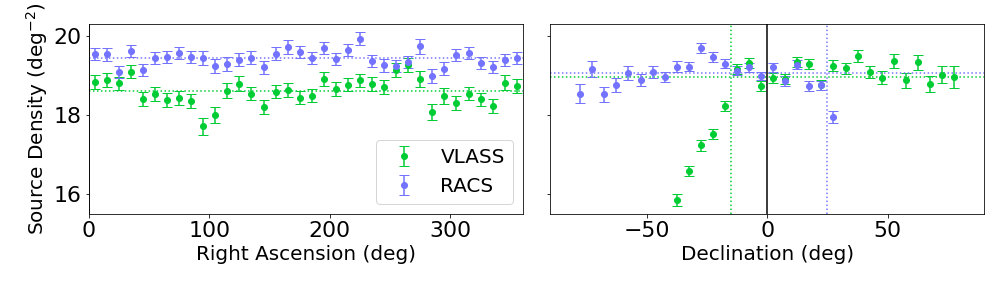}
\caption{
  Mean VLASS and RACS areal density versus right ascension (left) and declination (right) in $5^\circ$ increments.
  The vertical dotted lines denote the declination cuts used for the sinlge-survey dipole fitting ($\delta > -15^\circ$ for the VLASS
  and $\delta < +25^\circ$ for the RACS), and the horizontal dotted lines
  indicate the median values.  The vertical black
  line shows the declination where the two surveys are cut for the joint analysis.  The lower source density seen in the VLASS versus
  right ascension is caused by the falloff in source counts for $\delta < -15^\circ$.  When the indicated declination cut is applied, the
  source density in the VLASS matches that of the RACS.
}
\label{fig:source_density}
\end{centering}
\end{figure}

To assess the impact of combining two disparate radio surveys, we examine the source count dipole found in each survey alone.
As seen in Figure \ref{fig:surveys}, the surveys are not perfectly uniform, particularly as a function of declination.  
Figure \ref{fig:source_density} shows the areal source density versus equatorial coordinates.  The
VLASS shows a 16\% decrease from declination $-15^\circ$  to $-40^\circ$, and the RACS shows
a 6\% decrease for $\delta > +25^\circ$.   Neither survey shows trends in right ascension (the lowered source density in
the VLASS in Figure \ref{fig:source_density} (left) is caused by the diminished source density contribution from $\delta < -15^\circ$
regions, which affects all right ascension bins).

When we fit the VLASS or RACS alone without trimming the declination range, we obtain a dipole with apex near the apex of each survey (the equatorial poles)  because the reduced galaxy density at lower elevations (lower declinations for the VLASS, and higher declinations for
the RACS) produces a gradient.  Implementing a declination-based correction factor to source counts does not completely correct the
problem because there is spatial structure in the problematic declination ranges (Figure \ref{fig:surveys}).  

We also examined single-survey source count dipoles obtained by omitting the lower-density declinations indicated in Figure \ref{fig:source_density}.  If we trim the VLASS to the declinations that show no reduction in galaxy density, $\delta > -15^\circ$ and $f_{\rm sky} = 0.58$, then
the dipole nadir aligns with the missing data, directed toward the south equatorial pole: the apex is
$683^{+194}_{-113}$~km~s$^{-1}$ toward $93^{+8}_{-4}$, $56^{+20}_{-16}$ (degrees, Galactic coordinates).
However, using the RACS only we recover a galaxy count dipole that is consistent with the joint analysis, albeit with substantially larger uncertainties, which is a combination of fewer overall sources and significantly reduced sky coverage.  
% RACS-only SH fit is spot-on!
For the RACS catalog with $\delta < +25^\circ$ clipping, $f_{\rm sky} = 0.66$, and we obtain a dipole with amplitude 
%The RACS-only lmfit, on the other hand, is ok, with big errors:
$644^{+319}_{-148}$~km~s$^{-1}$ toward  $292^{+21}_{-23}$, $13^{+18}_{-20}$  (degrees).

If we fix the dipole direction to that obtained using the combined surveys and fit only for the amplitude, the VLASS, trimmed to
$\delta > -15^\circ$, has a reduced and less significant amplitude of $289^{+251}_{-192}$~km~s$^{-1}$ compared to the joint fit.
This is due in part to the low-count pixel regions seen near the dipole apex (Figure \ref{fig:surveys}, left).  For the RACS,
trimmed to $\delta < +25^\circ$, the amplitude is higher but has a similar uncertainty:   $428^{+203}_{-212}$~km~s$^{-1}$.
Both of these single-survey values are consistent with the CMB dipole amplitude.  
Given the amplitude uncertainties, however, this result cannot address the dipole amplitude frequency dependence
suggested by \citet{siewert2021}.

\section{Flux Limits}

The survey flux limits were selected to avoid the incompleteness turnover at low fluxes and the shot noise of rarer objects
at high fluxes.  \citet{blake2002} show a change in the observed radio galaxy dipole measurements with a change in lower flux
limit, and we assess the impact of the lower flux limit by doubling the lower flux limit on the VLASS, propagating it to RACS,
and following the procedures described in Sections \ref{sec:data} and \ref{sec:meas}.  Doubling the lower flux limit roughly halves the
source catalog to 54\% or the original catalog
(because the exponent on $N(>S)\propto S^{-x}$ is $x=-1$), correspondingly lowering the signal-to-noise.
With the reduced catalog, we recover the source count dipole, but with larger uncertainties (as expected):
$657^{+429}_{-135}$~km~s$^{-1}$ toward $309^{+24}_{-44}$, $27^{+20}_{-27}$ (degrees).

\section{Peak versus Total Flux}

To explore the choice of using the peak flux density from the VLASS and RACS surveys rather than the more canonical total source flux density, we repeated the analysis using the total source flux in both catalogs.  For simplicity, we retain the flux limits used for the VLASS
and scale these to the RACS using the median spectral index obtained from the total flux ratio in the overlap region (Section \ref{sec:data}).
In this case, $\alpha = 0.81$, the count index $x$ is unchanged, and we recover the dipole:  $527^{+195}_{-132}$~km~s$^{-1}$ toward $241^{+17}_{-49}$, $66^{+10}_{-19}$ (degrees),  which is consistent with the
  CMB vector (a $\chi^2_3$ test $p$-value of 0.25).
For the reasons described in Section \ref{sec:data}, the peak flux densities are preferable, but the total flux densities can also be used to detect a galaxy dipole that is consistent with the CMB.

\section{Radio Spectral Index}

The fine-tuning of the spectral index $\alpha$ used to combine the surveys is a concern because the spectral index shows
a large range of values among the VLASS-RACS overlap sample (Section \ref{sec:data}).  In our treatment, we employ the median
spectral index or the flux-weighted median spectral index of the
overlap sample.  When the spectral index is incorrect, a dipole is produced that points toward the equatorial hemisphere favored by the
over-weighted survey.  Fine turning is required to equalize the two hemispheres to less than 1\% in order to detect the dipole
signal.

In order to address this fine-tuning problem and the large intrinsic dispersion in spectral indices, we use the overlap region between
the surveys ($|\delta| < 10^\circ$) to find the RACS flux limits that produce a match between the areal source densities of the two surveys given the VLASS flux
limits.  The upper RACS flux limit has negligible impact on the source counts.  Adjusting the lower RACS flux limit to obtain matching
source count densities in the overlap region between the two surveys produces a limit that agrees with the flux limit obtained from the
median spectral index to within 1\%.  It is notable that the combined overlap catalog obtained from matching VLASS to RACS sources
(Section \ref{subsec:matching}) is not identical to the individual flux-limited VLASS and RACS catalogs in the overlap region.  It is
reassuring that the selection outcomes agree between these disparate approaches to catalog matching.

Ultimately, choosing a median spectral index and the above count-matching approach are equivalent, but the latter may be
more satisfying from a fine-turning perspective.  The count-matching approach could be modified to a flux-matching approach
to measure the radio flux dipole.

\end{document}